\documentclass[aip]{revtex4-1}   
\usepackage{amsmath,xcolor}\def\d{{\, \rm d}}
\usepackage{graphicx}
\graphicspath{ {./figures/} }
\usepackage{verbatim}
\begin{document}


\title{An Efficient Continuous Data Assimilation Algorithm for the Sabra Shell Model of Turbulence} 

\author{Nan Chen}
\email[]{chennan@math.wisc.edu}
\affiliation{Department of Mathematics, University of Wisconsin-Madison, Madison, WI 53706, USA}

\author{Yuchen Li}
\email[]{yli966@wisc.edu}
\affiliation{Department of Mathematics, University of Wisconsin-Madison, Madison, WI 53706, USA}

\author{Evelyn Lunasin}
\email[]{lunasin@usna.edu}
\affiliation{Department of Mathematics, United States Naval Academy, Annapolis, MD 21402, USA}


\date{\today}

\begin{abstract}
Complex nonlinear turbulent dynamical systems are ubiquitous in many areas. Recovering unobserved state variables is an important topic for the data assimilation of turbulent systems. In this article, an efficient continuous in time data assimilation scheme is developed, which exploits closed analytic formulae for updating the unobserved state variables. Therefore, it is computationally efficient and accurate. The new data assimilation scheme is combined with a simple reduced order modeling technique that involves a cheap closure approximation and a noise inflation. In such a way, many complicated turbulent dynamical systems can satisfy the requirements of the mathematical structures for the proposed efficient data assimilation scheme. The new data assimilation scheme is then applied to the Sabra shell model, which is a conceptual model for nonlinear turbulence. The goal is to recover the unobserved shell velocities across different spatial scales. It has been shown that the new data assimilation scheme is skillful in capturing the nonlinear features of turbulence including the intermittency and extreme events in both the chaotic and the turbulent dynamical regimes. It has also been shown  that the new data assimilation scheme is more accurate and computationally cheaper than the standard ensemble Kalman filter and nudging data assimilation schemes for assimilating the Sabra shell model.
\end{abstract}


\maketitle 

\section{Introduction}
Complex anisotropic turbulent systems are ubiquitous in geophysics, engineering and climate science \cite{majda2016introduction, strogatz2018nonlinear, sheard2009principles}. These systems often contain a high-dimensional phase space with strong nonlinear interactions between state variables across different temporal and spatial scales. Intermittency, extreme events and regime switching are some of the typical features in these nonlinear turbulent systems \cite{beniston2007future, mohamad2015probabilistic, franzke2012predictability}. One major challenge in studying these complex nonlinear systems is the availability of only partial observations in practice. Developing efficient data assimilation algorithms thus becomes necessary for the state estimation of unobserved variables \cite{majda2018model, gershgorin2011filtering, law2015data}.

Due to the improvement of satellites and other refined observational networks, the available observations nowadays are quite frequently related to the time scale of the quantities of interest in many applications. Therefore, the recovery of the unobserved state variables in these situations can be treated as a continuous in time data assimilation problem, which is the focus of this article.
Nudging (also known as Newtonian Relaxation) is one of the most widely used continuous in time data assimilation methods \cite{hoke1976initialization, stauffer1990use}. It relaxes the model state towards the observations by the addition of terms to the original differential equations, which are proportional to the difference between the observations and the model states.  Compared with other data assimilation approaches, nudging is computationally efficient \cite{FoiasManleyRosaTemamBook2001}, and it has been used in many practical applications \cite{stauffer1994multiscale, leidner2001improving, duane2006synchronicity, dixon2009impact}. The simple mathematical structure of the nudging technique also facilitates using rigorous mathematical analysis to understand its data assimilation skill \cite{AzouaniOlsenTiti, FarhatLunasinTiti2017a, FarhatLunasinTiti2017b}. However, one major challenge in applying nudging is determining the nudging coefficients, which in practice are typically carried out by ad hoc tuning. Therefore, the optimality of the state estimation from nudging is often not guaranteed. Note that the nudging with such constant coefficients is conceptually equivalent to utilizing a fixed ``Kalman gain'' matrix in the nonlinear version of the Kalman-Bucy filter, which means the background covariance matrix is pre-determined. Thus, the state estimation using nudging often suffers from large errors in the presence of partial observations for complex nonlinear systems with strong turbulence, intermittency and extreme events.
On the other hand, the ensemble Kalman filter (EnKF) \cite{law2015data, evensen2003ensemble, evensen2009data, burgers1998analysis}, which utilizes the Bayesian formula that combines model forecast with observations to seek the optimal state estimation, is more skillful in dealing with strong turbulence and nonlinearity. Importantly, it also quantifies the time evolution of the uncertainty in the state estimation. The EnKF has led to many successes in assimilating complex nonlinear turbulent systems \cite{evensen1996assimilation, massonnet2014calibration, houtekamer2005atmospheric}. One potential issue in applying the EnKF is the sampling error due to the fact that only a small number of ensembles is affordable for assimilating high-dimensional systems. Therefore, many empirical tuning techniques \cite{law2015data, evensen2003ensemble}, including noise inflation, localization and resampling, are essential to prevent the filter divergence and improve the accuracy of the results in practice. Comparing with the straightforward nudging data assimilation, the EnKF is more computationally demanding. In particular, the computational efficiency decreases significantly in the presence of dense or nearly continuous in time observations, which require a large number of assimilation cycles.

In this article, an efficient continuous data assimilation algorithm is developed for estimating the unobserved state and the associated uncertainty, which is then applied to recovering the unobserved states in the Sabra shell models given partial observations. This new continuous in time data assimilation scheme exploits closed analytic formulae to recover the unobserved states. Therefore, the state estimation is exact and accurate, which avoids the sampling error and the associated tuning resulting from ensemble or Monte Carlo simulations. The closed analytic formulae also allows the new scheme to be computationally efficient, which is then directly applicable to high-dimensional systems. On the other hand, this new data assimilation scheme provides the optimal estimate of the time evolution of the uncertainty in addition to the optimal point-wise mean state estimation, which is fundamentally different from the nudging data assimilation. Including such an uncertainty is crucial in assimilating complex systems with strong nonlinear and turbulence. In addition, this new efficient data assimilation algorithm can be effectively combined with several reduced order modeling approaches for further reducing the computational cost as well as the model error in assimilating complex nonlinear system in practice.

Shell models are simple nonlinear systems of complex ordinary differential equations that mimic the scale-by-scale local transfer of energy and the overall energy cascade of three-dimensional (3D) turbulence but with a much reduced degrees of freedom \cite{lorenz1972, Siggia78, Gledzer73, OhkitaniYamada89, Lvov98, Ditlevsen2000}.  The Sabra shell model of turbulence is an improved shell model  possessing  an additional structure on the nonlinear terms which prevents large oscillations in the scaling regime and avoids slowly decaying correlations between velocity components with different wave numbers\cite{Lvov98}, making  numerical simulation less tedious than its counterparts.  It is constrained by the conservation of the global "energy" and "helicity" (in the absence of viscosity and body force).    It exhibits similar anomalies on the scaling exponents and spectral scaling behavior in the inertial range similar to what has been observed in 3D fluid flow experiments \cite{Biferale2003}. Therefore, the Sabra shell model is a widely adopted test model for nonlinear turbulent systems.

The rest of the article is organized as follows. Section \ref{Sec:DA_Method} presents the new continuous in time data assimilation framework. The Sabra shell model and its features analogous to turbulent flows are summarized in Section \ref{Sec:Sabra}. The data assimilation of the Sabra shell model is studied in Section \ref{Sec:DA}. The article is concluded in Section \ref{Sec:Conclusion}.

\section{The New Efficient Continuous in Time Data Assimilation Algorithm}\label{Sec:DA_Method}
\subsection{The general framework}
The new continuous in time data assimilation algorithm is applicable to a rich class of nonlinear turbulent systems. Consider the following coupled nonlinear model,
\begin{subequations}\label{General_CG}
\begin{align}
 \frac{\d\mathbf{v}}{\d t} &= \mathbf{A}_{\mathbf{0}}(\mathbf{v},t) + \mathbf{A}_{\mathbf{1}}(\mathbf{v},t) \mathbf{w} + \boldsymbol\sigma_v(\mathbf{v},t) \dot{\mathbf{W}}_v, \label{General_CG_v}\\
 \frac{\d\mathbf{w}}{\d t} &= \mathbf{a}_{\mathbf{0}}(\mathbf{v},t) + \mathbf{a}_{\mathbf{1}}(\mathbf{v},t) \mathbf{w} + \boldsymbol\sigma_w(\mathbf{v},t) \dot{\mathbf{W}}_w, \label{General_CG_w}
\end{align}
\end{subequations}
where $\mathbf{v}$ and $\mathbf{w}$ are multidimensional state variables. In \eqref{General_CG}, $\mathbf{A}_{\mathbf{0}}$ and $\mathbf{a}_{\mathbf{0}}$ are vectors while $\mathbf{A}_{\mathbf{1}}$ and $\mathbf{a}_{\mathbf{1}}$ are matrices, all of which are any nonlinear functions of the state variable $\mathbf{v}$ and time $t$. The terms $\dot{\mathbf{W}}_v$ and $\dot{\mathbf{W}}_w$ are independent white noise, where the noise coefficients $\boldsymbol\sigma_v$ and $\boldsymbol\sigma_w$ can also be functions of the state variable $\mathbf{v}$ and time $t$. The coupled model \eqref{General_CG} is apparently nonlinear and it has non-Gaussian statistics. With some minor approximations, many complex turbulent partial differential equations (PDEs), such as the Navier-Stokes equations (NSE), can be written into the form of \eqref{General_CG} \cite{chen2018conditional, chen2016filtering, chen2016model} (see Section \ref{Sec:ROM}).

One desirable feature of the coupled system \eqref{General_CG} is its conditional Gaussianity. Specifically, conditioned on one realization (i.e., a random trajectory) of $\mathbf{v}$, the conditional distribution for the hidden variable $\mathbf{w}$ at time $t$ conditioned on observations up to time $t$,
\begin{equation}\label{CG_PDF}
p(\mathbf{w}(t)|\mathbf{v}(s\leq t)\big)\sim\mathcal{N}(\boldsymbol\mu,\mathbf{R})
\end{equation}
 is Gaussian, with mean ${\boldsymbol \mu}$ and covariance $\mathbf{R}$.  This is because with a given trajectory $\mathbf{v}(s\leq t)$, the system \eqref{General_CG} is a linear system in terms of $\mathbf{w}$ and the noise coefficients do not depend on $\mathbf{w}$, either. Such a conditional Gaussian distribution is the so-called posterior distribution in  data assimilation. This can be seen by regarding \eqref{General_CG_w} as the forecast model of $\mathbf{w}$ while the process of $\mathbf{v}$ in \eqref{General_CG_v} represents a continuous nonlinear observation of $\mathbf{w}$. In the special case that all the matrices and vectors on the right hand side of \eqref{General_CG} have no dependence on the state variable $\mathbf{v}$, the conditional distribution \eqref{CG_PDF} is the solution of the associated Kalman-Bucy filter \cite{kalman1961new}.

Despite the nonlinearity in both the forecast model and observational process from \eqref{General_CG}, the conditional distribution  \eqref{CG_PDF} can be expressed by the following closed analytic formulae \cite{liptser2001statistics},\begin{subequations}\label{CGNS_Stat}
\begin{align}
  \d \boldsymbol{\mu} &= (\mathbf{a}_\mathbf{0} + \mathbf{a}_\mathbf{1} \boldsymbol{\mu})\d t + \mathbf{R}\mathbf{A}_\mathbf{1}^*  (\boldsymbol\sigma_v\boldsymbol\sigma_v^*)^{-1}(\d\mathbf{v} - (\mathbf{A}_\mathbf{0} + \mathbf{A}_\mathbf{1}\boldsymbol{\mu})\d t),\label{CGNS_Stat_Mean}\\
  \d\mathbf{R} &= \big(\mathbf{a}_\mathbf{1} \mathbf{R} + \mathbf{R}\mathbf{a}_\mathbf{1}^* + \boldsymbol\sigma_w\boldsymbol\sigma_w^*-  \mathbf{R}\mathbf{A}_\mathbf{1}^*(\boldsymbol\sigma_v\boldsymbol\sigma_v^*)^{-1} \mathbf{A}_\mathbf{1}\mathbf{R}\big)\d t,\label{CGNS_Stat_Cov}
\end{align}
\end{subequations}
where the asterisk stands for conjugate transpose. The derivation of the formulae in \eqref{CGNS_Stat} follows the Bayesian inference \cite{liptser2001statistics}, where the one-step forecast from \eqref{General_CG_w} is regarded as the prior while the equation \eqref{General_CG_v} provides the likelihood.
One crucial feature to highlight is that the exact and accurate  data assimilation scheme \eqref{CGNS_Stat} avoids the sampling error and the associated tuning resulting from ensemble or Monte Carlo simulations. The scheme in \eqref{CGNS_Stat} also allows a low computational cost and is then applicable to high-dimensional systems.
It is important to note that, since $\mathbf{A}_\mathbf{1}$ is a function of $\mathbf{v}$, the time evolution of the posterior uncertainty $\mathbf{R}$ will not converge to a constant matrix. This is fundamentally different from the Kalman-Bucy filter and the nudging data assimilation. It is this non-stationary feature in $\mathbf{R}$ that facilitates the skillful recovery of the nonlinear and turbulence features, including the intermittency and extreme events.

\subsection{Incorporating complex nonlinear PDE systems into the new data assimilation framework}\label{Sec:ROM}
To incorporate a complex nonlinear PDE system $\mathcal{M}$ into the new nonlinear modeling framework \eqref{General_CG}, the first step is to find a set of basis functions $\{\boldsymbol\varphi_1,\boldsymbol\varphi_2,\ldots,\boldsymbol\varphi_R\}$. Projecting $\mathcal{M}$ onto these basis leads to
\begin{equation}\label{Projected_Model}
  \frac{\d\mathbf{u}_R}{\d t} =  \mathbf{A}_u\mathbf{u}_R + \mathbf{u}_R^*\mathbf{B}_u\mathbf{u}_R + \mbox{residual},
\end{equation}
where $\mathbf{u}_R = (\hat{{u}}_1,\ldots,\hat{{u}}_R)^{\mathtt{T}}$ is the collection of the state variables in the projection space with $\mathbf{u}_R = \sum_{i=1}^R \hat{{u}}_i\boldsymbol\varphi_i$ being the approximate solution of $\mathcal{M}$. In \eqref{Projected_Model}, the notation $\mathbf{u}_R^*$ represents the conjugate transpose of $\mathbf{u}_R$ while $\mathbf{A}_u$  and $\mathbf{B}_u$ are matrices that do not dependent on $\mathbf{u}_R$. Here, for the conciseness of presentation, we have implicitly assumed that the nonlinearity in the starting PDE system $\mathcal{M}$ is quadratic and therefore only the linear term $\mathbf{A}_u\mathbf{u}_R$ and the quadratic nonlinear term $\mathbf{u}_R^*\mathbf{B}_u\mathbf{u}_R$ appear in the projected system \eqref{Projected_Model}. The starting PDE system with higher order nonlinearity can be easily dealt with in a similar way. The last term on the right hand side of \eqref{Projected_Model} represents the residual due to the Galerkin truncation of $\mathcal{M}$.

Next, the $R$ modes of $\mathbf{u}_R$ are categorized into two groups: $\mathbf{v}$ and $\mathbf{w}$, which represent the observed and unobserved variables, respectively. Therefore, the system \eqref{Projected_Model} can be rewritten as
\begin{subequations}\label{Decomposed_Model}
\begin{align}
 \frac{d\mathbf{v}}{dt} &= \mathbf{A}^{(v)}_{v}\mathbf{v} + \mathbf{v}^*{\mathbf{B}}^{(v)}_{vv}\mathbf{v} + \mathbf{v}^*{\mathbf{B}}^{(v)}_{vw}\mathbf{w} + \mathbf{w}^*{\mathbf{B}}^{(v)}_{ww}\mathbf{w} + \mbox{residual}_1, \\
 \frac{d\mathbf{w}}{dt} &= \mathbf{A}^{(w)}_{w}\mathbf{w} + \mathbf{v}^*{\mathbf{B}}^{(w)}_{vv}\mathbf{v} + \mathbf{v}^*{\mathbf{B}}^{(w)}_{vw}\mathbf{w} + \mathbf{w}^*{\mathbf{B}}^{(w)}_{ww}\mathbf{w} + \mbox{residual}_2,
\end{align}
\end{subequations}
where the ${\mathbf{B}}^{(\cdot)}_{\cdot}$ in \eqref{Decomposed_Model} are related to the $\mathbf{B}_u$ in \eqref{Projected_Model}.

Finally, to formulate \eqref{Decomposed_Model} into the form in \eqref{General_CG}, the self-interaction of $\mathbf{w}$, namely the quadratic nonlinearity between the unobserved variables themselves, together with the residual terms are dropped. To compensate, additional parameterizations and stochastic noise are utilized to approximate the contribution from these terms, namely,
\begin{subequations}\label{Parameterization_Terms}
\begin{align}
  \mathbf{w}^*{\mathbf{B}}^{(v)}_{ww}\mathbf{w} + \mbox{residual}_1 &\approx  \boldsymbol\tau^{(v)}_{v}(\mathbf{v}) + \boldsymbol\sigma_v \dot{\mathbf{W}}_v,\label{Parameterization_Terms_v}\\
  \mathbf{w}^*{\mathbf{B}}^{(w)}_{ww}\mathbf{w} + \mbox{residual}_2 &\approx  \boldsymbol\tau^{(w)}_{v}(\mathbf{v}) + \boldsymbol\sigma_w \dot{\mathbf{W}}_w.\label{Parameterization_Terms_w}
\end{align}
\end{subequations}
One of the motivations of approximating the quadratic nonlinearity between the unobserved variables themselves is the following. In many applications, these terms represent the self-interactions between high frequencies. Therefore, stochastic noise is a suitable surrogate to approximate these fast variabilities, which has been justified in the MTV (\underline{M}ajda-\underline{T}imofeyev-\underline{V}anden Eijnden) strategy \cite{majda2001mathematical, majda1999models}. Yet, the unobserved variables in general may not contain only the fast components. Therefore, it is essential to further include additional deterministic parameterizations that describe the slow-varying components in the self-interactions between the hidden variables. A simple but effective approach is to incorporate the closure terms $\boldsymbol\tau^{(v)}_{v}(\mathbf{v})$ and $\boldsymbol\tau^{(w)}_{v}(\mathbf{v})$ that depend only on the observed variables $\mathbf{v}$ \cite{xie2018data, mou2020data, san2018extreme}, which allows the approximate model to have the desired mathematical structure as  the system \eqref{General_CG}.
Collecting all the above information yields the following coupled model,
\begin{subequations}\label{ROM_CG}
\begin{align}
 \frac{d\mathbf{v}}{dt} &= \mathbf{A}^{(v)}_{v}\mathbf{v} + \mathbf{v}^*\mathbf{B}^{(v)}_{vv}\mathbf{v} + \mathbf{v}^*\mathbf{B}^{(v)}_{vw}\mathbf{w} + \boldsymbol\tau^{(v)}_{v}(\mathbf{v}) + \boldsymbol\sigma_v \dot{\mathbf{W}}_v,\label{ROM_CG_a}\\
 \frac{d\mathbf{w}}{dt} &= \mathbf{A}^{(w)}_{w}\mathbf{w} + \mathbf{v}^*\mathbf{B}^{(w)}_{vv}\mathbf{v} + \mathbf{v}^*\mathbf{B}^{(w)}_{vw}\mathbf{w} + \boldsymbol\tau^{(w)}_{v}(\mathbf{v}) + \boldsymbol\sigma_w \dot{\mathbf{W}}_w.\label{ROM_CG_b}
\end{align}
\end{subequations}
Note that the stochastic noise can also be utilized to compensate for the model error, since the perfect model is never known in practice. For the convenience of presentation, we name the new efficient continuous in time data assimilation scheme \eqref{CGNS_Stat} associated with the system \eqref{ROM_CG} as a conditional Gaussian nonlinear data assimilation (CGNDA) since the conditional statistics in \eqref{CGNS_Stat} is Gaussian while the coupled system in \eqref{ROM_CG} is nonlinear.

As a final remark, if the coupled system \eqref{ROM_CG} is utilized for medium- or long-range forecast, then it is more appropriate to slightly adjust the nonlinear terms on the right hand side of \eqref{ROM_CG} such that the physics constraints (i.e., the conservation of energy in the nonlinear terms) \cite{majda2012physics, harlim2014ensemble} are satisfied. The physics constraints have been shown to be crucial in preventing the finite-time blow-up of the solution \cite{majda2012fundamental}. Nevertheless, for the continuous in time data assimilation, the physics constraint is not as crucial as the long-range forecast since the observations continuously correct the model forecast bias.
Similarly, to better approximate the self-interactions between the hidden variables $\mathbf{w}$ for the purpose of forecast,  a second nonlinear parameterization term involving a conditional linear component of $\mathbf{w}$ itself can be included in each equation in \eqref{Parameterization_Terms} (namely adding $\boldsymbol\tau^{(v)}_{w}(\mathbf{v})\mathbf{w}$ to \eqref{Parameterization_Terms_v} and $\boldsymbol\tau^{(w)}_{w}(\mathbf{v})\mathbf{w}$ to \eqref{Parameterization_Terms_w}). The resulting system with such a refined parameterization  preserves the mathematical structures in \eqref{General_CG},
\begin{subequations}\label{ROM_CG2}
\begin{align}
 \frac{d\mathbf{v}}{dt} &= \mathbf{A}^{(v)}_{v}\mathbf{v} + \mathbf{v}^*\mathbf{B}^{(v)}_{vv}\mathbf{v} + \mathbf{v}^*\mathbf{B}^{(v)}_{vw}\mathbf{w} + (\boldsymbol\tau^{(v)}_{w}(\mathbf{v})\mathbf{w} + \boldsymbol\tau^{(v)}_{v}(\mathbf{v})) + \boldsymbol\sigma_v \dot{\mathbf{W}}_v,\label{ROM_CG2_a}\\
 \frac{d\mathbf{w}}{dt} &= \mathbf{A}^{(w)}_{w}\mathbf{w} + \mathbf{v}^*\mathbf{B}^{(w)}_{vv}\mathbf{v} + \mathbf{v}^*\mathbf{B}^{(w)}_{vw}\mathbf{w} + (\boldsymbol\tau^{(w)}_{w}(\mathbf{v})\mathbf{w} + \boldsymbol\tau^{(w)}_{v}(\mathbf{v})) + \boldsymbol\sigma_w \dot{\mathbf{W}}_w.\label{ROM_CG2_b}
\end{align}
\end{subequations}

\section{The Sabra Shell Model}\label{Sec:Sabra}
Shell models are phenomenological models derived to have several structurally similar spectral properties as that of  Navier-Stokes equations (NSE) but with highly reduced degrees of freedom \cite{lorenz1972, Siggia78, Gledzer73, OhkitaniYamada89, Lvov98, Ditlevsen2000}. These models are of practical interest because of their ability to numerically reproduce intermittencies with high-order structure functions scaling exponents consistent with those in the 3D turbulent flow as are observed in experiments \cite{Biferale2003}.  The shell models capture these statistics under high computational efficiency  by retaining only very few representative variables for each shell. The $n$-th shell contains the information within the octave of wave numbers $\lambda^n \leq |\tilde{{k}}_n| < \lambda^{n+1}$, where $\tilde{{k}}_n$ is the complex Fourier wavenumber of the original 3D turbulent model and $\lambda>1$ is a pre-determined constant. Define $k_n = k_0\lambda^n$ as an analog of the Fourier wavenumber in the shell model, where the prefactor $k_0$ is the inverse of the integral scale $L_0 = k_0^{-1}$.
The shell model then characterizes the time evolution of the ``averaged'' velocities in each shell.
The complex shell velocity $u_n$ represents the velocity fluctuations corresponding to the $n$-th shell for $n = 1, \ldots, N$.  With this representation,  there are only  $\sim\log k_N$ state variables to be computed in the shell model in order to include the information up to Fourier wave number $k_0\lambda^N$ in the original 3D system. Nevertheless, the shell model is able to simulate very high Reynolds number flows.   It is also worthwhile to mention that shell velocities being defined to be complex variables are observed to visit all parts of the attractor more rapidly \cite{Ditlevsen2000}.

The general form of the shell models is a nonlinear system of ordinary differential equations,
\begin{equation}\label{shell}
\dfrac{d}{dt} u_n + \nu k_n^2 u_n = k_n S_n[\mathbf{u},\mathbf{u}] + f_n,\qquad n = 1,\ldots, N,
\end{equation}
where $\mathbf{u}$ is a complex vector that contains all the complex state variables $u_n$. The  nonlinear terms $S_n[\mathbf{u},\mathbf{u}]$ are structured to preserve certain ideal invariants and mimic local triad interactions analogous to that of the original 3D NSE.  The velocity fluctuations are assumed to be zero  on all scales larger than $L_0$. Similarly, the terms $\nu k_n^2 u_n$ and $f_n$ are the analogs to the (Fourier transformed)  viscous term and external forcing in the NSE, respectively, where the parameter $\nu>0$ is the "viscosity" and the external forcing in the shell models is typically restricted to only the first few shells that represent the large-scale variabilities.

The Sabra shell model \cite{Lvov98} has specific interaction coefficients on the nonlinear terms,
\begin{equation}\label{sabra}
\frac{d u_n}{dt} + \nu k_n^2 u_n
 = i (ak_{n+1} u^*_{n+1} u_{n+2} +b k_n u^*_{n-1} u_{n+1}  -c k_{n-1} u_{n-1} u_{n-2} ) + f_n,\qquad n = 1\dots N,
 \end{equation}
where $\cdot^*$ denotes the complex conjugation and the real-valued parameters $a,b,c$ are chosen such that $a+b+c=0$.  The boundary conditions are given with the assignment $u_{-1} = u_0$   and  $u_{N+1} = u_{N+2} =0$. Note that  in the absence of viscosity  and external forcing, the quantities
\begin{equation*}
E=\sum_n |u_n|^2 \qquad\mbox{and}\qquad H=\sum_n \left(\frac{a}{c}\right)^n  |u_n|^2
\end{equation*}
are conserved.  These quantities are the analogues of the energy and the helicity, respectively, which are the ideal invariants in the 3D NSE.

\section{Data Assimilation of the Sabra Shell Model}\label{Sec:DA}

\subsection{Setup}
\subsubsection{Model parameters}
The Sabra shell model \eqref{sabra} is simulated with the intershell ratio given by $\lambda = 2$, so that $k_n = k_0\lambda^n$, with $k_0 = 2^{-4}$.  The interaction coefficients are set to  $a = 1, b = c =  -1/2$. Constant forcing with magnitude one are imposed onto the first two shells $n=1$ and $n=2$, which represent the large-scale background mean flow \cite{Levant2010}.

\subsubsection{Observed and unobserved variables}
In the following data assimilation experiments, assume the time series of the complex shell velocities $u_1, u_2, u_5$ and $u_6$ are observed in a continuous-in-time fashion. The goal is to recover the unobserved complex shell velocities $u_3, u_4, u_7$ and $u_8$. The shell velocities $u_1$ and $u_2$ represent the dynamics on the largest scale while $u_5$ and $u_6$ mimic the medium-scale variabilities. In a typical scenario, the information of the largest scale is available from satellite observations while the variabilities within the medium-range spatial scales are possibly obtained by local observations, such as tracers or drifters in the ocean. It is important to note that the scale gap between $u_2$ and $u_5$ associated with the original turbulent flow is $\lambda^3 = 8$. Therefore, there is a wide range of the scales within which the recovery of the dynamical features is required. These scales correspond to the shell velocities $u_3$ and $u_4$. On the other hand, the shells $u_7$ and $u_8$ in the Sabra shell model correspond to the small scales that are typically unresolved but remain a direct interaction with the resolved scale variables, which can be seen from the triad interactions in the Sabra model \eqref{sabra}. Such a setup, corresponding to the CGNDA framework \eqref{ROM_CG}, implies
\begin{equation}\label{Sabra_DA_modes}
\begin{split}
\mathbf{v} &= (u_1, u_2, u_5, u_6)\\
\mathbf{w} &= (u_3, u_4, u_7, u_8).
\end{split}
\end{equation}
Note that the model that generates the true signal is the original Sabra model \eqref{sabra}, which contains more than the $8$ shells described in \eqref{Sabra_DA_modes}. The solution from \eqref{sabra} is named as the reference solution.
\subsubsection{Dynamical regimes}
\begin{figure}
\includegraphics[width=16cm]{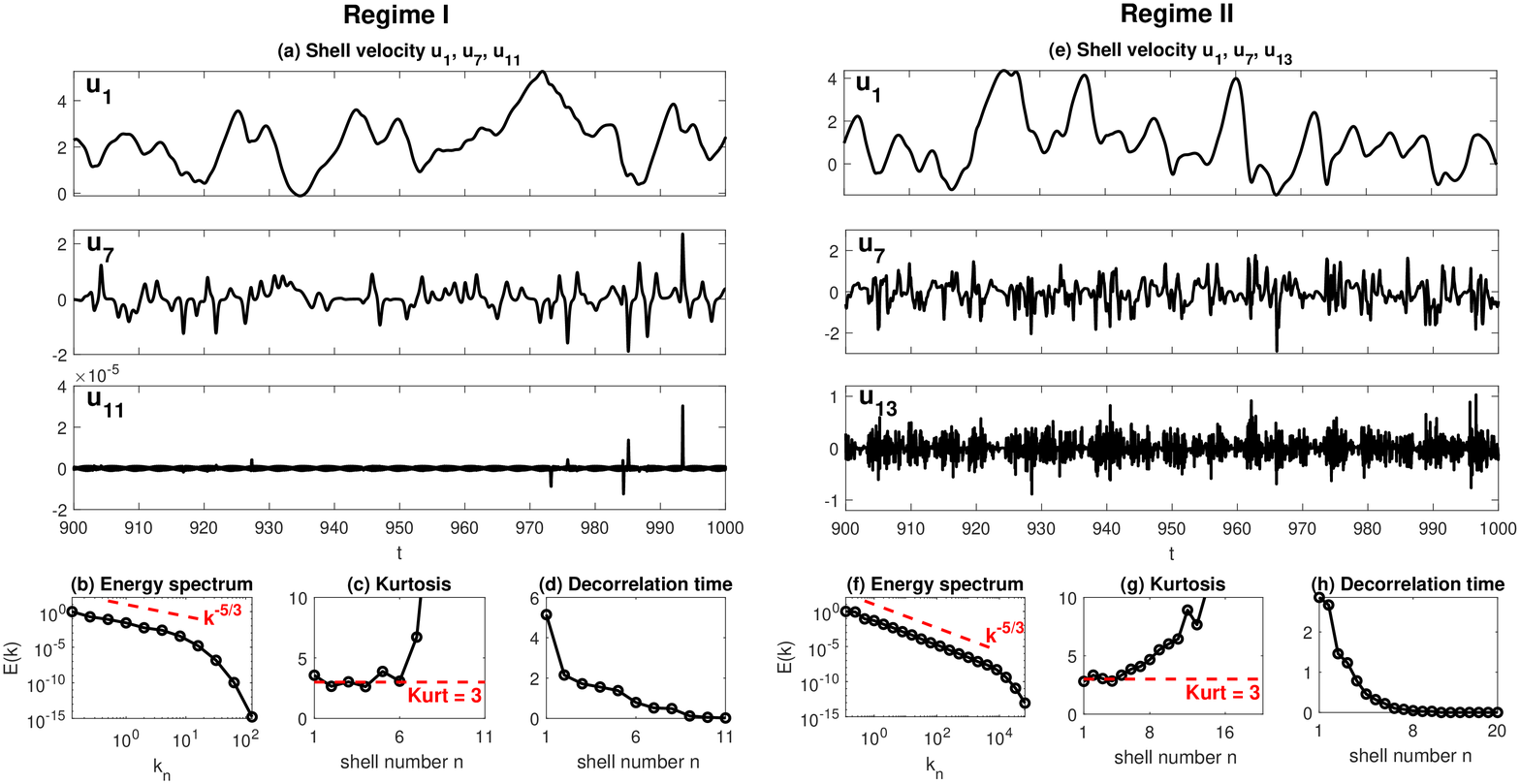}
\caption{Two dynamical regimes of the Sabra shell model \eqref{sabra}. In Regime I (Panels (a)--(d)), a moderate viscosity $\nu = 0.09$ is taken and the total number of the shells is $N=11$. In Regime II (Panels (e)--(h)), a tiny viscosity $\nu = 10^{-5}$ is adopted and the total number of the shells is $N=20$. For both regimes, the time series of three different shells, the energy spectrum, the kurtosis and the decorrelation time of the time series associated with each shell velocity are shown. The red dashed line in Panels (b) and (f) show the Kolmogorov spectral scaling $k^{-5/3}$ and those in Panels (c) and (g) show the kurtosis $= 3$ value, beyond which the PDF has fat tails. }\label{Dynamical_Regimes}
\end{figure}

Two dynamical regimes, which are differed by the viscosity coefficient $\nu$, are studied here.
Regime I corresponds to a moderate viscosity $\nu = 0.09$ with a total number of the shells being $N=11$. Regime II involves a tiny viscosity $\nu = 10^{-5}$ and the total number of the shells is $N=20$. Figure \ref{Dynamical_Regimes} shows the key features of the two dynamical regimes, including the time series, the energy spectrum, the kurtosis  and the decorrelation time of the time series associated with each shell velocity. The energy spectrum is given by the following formula, $E(k_n) = {e(k_n)}/{k_n}$, where $e(k_n)$ is the energy of the time series $u_k$.  In Regime I, the inertial range is up to $n=6$, which satisfies the Kolmogorov spectral scaling $E(k_n)\sim k^{-5/3}$. The energy in the shells starts to decrease more rapidly at $n=7$. Meanwhile, the non-Gaussian features become significant with kurtosis being much larger than $3$. In contrast, the inertial range of Regime II contains many more wavenumbers. The non-Gaussian features in the shell velocities become obvious for shell numbers that are greater than $n=5$. These non-Gaussian features, including the intermittency and extreme events, are also clearly illustrated in the time series. 

\subsubsection{Numerical schemes}
The results from the CGNDA are compared with those from the EnKF and the nudging data assimilation. The perfect model, with $N=11$ and $N=20$ for Regime I and Regime II, respectively, is used as the forecast model for the EnKF and the nudging data assimilation schemes. The ensemble transform Kalman filter (ETKF) \cite{bishop2001adaptive} is utilized as a robust scheme for the EnKF. The true Sabra model and the nudging data assimilation are both solved via an adaptive fourth order Runge-Kutta algorithm. Only a forward Euler-Maruyama scheme with a small time step $\Delta{t}=10^{-3}$ is used for solving the CGNDA. The same $\Delta{t}$ is used as the numerical discrete approximation for the continuous observations. For the ETKF,  a forward Euler-Maruyama scheme with a time step $\Delta{t}=10^{-4}$ is used to resolve the smaller scale variables. A tiny observational noise is added onto the observations of the ETKF to prevent the singularity of the scheme, which itself will not cause any significant decrease of the data assimilation skill. The ensemble size of the ETKF is $200$ and $800$ for Regime I and Regime II, respectively, which has been tested to be large enough that eliminates the sampling error. It is important to note that running the ETKF (or in general EnKF) is much more expensive than the CGNDA. For the nudging data assimilation, the approximate solution ${\bf U}$ of ${\bf u}$  evolves as follows,
    \begin{equation}\label{feedback:sys}
        \frac{\d\mathbf{U}}{\d t}=\mathbf{F}(\mathbf{U})
        -\mu \delta_{n,n_o} ( {\bf U } - \mathcal{H}{\bf u} ), \quad {\bf U}(0) = {\bf 0},
    \end{equation}
where $\mathbf{F}$ describes the dynamics of the governing model, $\delta_{n,n_o}$ denotes the standard Kronecker delta function which equals to 1 when the complex components of ${\bf U}$ indexed by $n = 1\dots N$ take on the index values  $n_0 = 1,2,5,6$ and the corresponding observation operator defined by a diagonal matrix $ \mathcal{H} $ which consists of 1 on the diagonal entries on the indices $(n_0,n_0)$ corresponding to the observational state variables.   The zero initial condition is chosen for simplicity.   The observational data is inserted into the model via the feedback term with the relaxation parameter $\mu>0$  chosen large enough to drive the time series of the approximate solution to the  observations but not too large that the induced oscillations in the small scales cannot be suppressed by the damping term which corresponds to the viscous dissipation in the original NSE. A range of values is known to provide such a case and one can obtain a semi-optimal value by computing the correlation of the time series of approximate solution and the reference solution. Here $\mu=2$ is utilized.

\subsection{Determining the parameterization and noise coefficients in the CGNDA model}\label{Sec:coefficients}
Recall that in \eqref{Parameterization_Terms}, the two parameterized terms $\boldsymbol\tau^{(v)}_{v}(\mathbf{v})$ and $\boldsymbol\tau^{(w)}_{v}(\mathbf{v})$ and the two noise coefficients $\boldsymbol\sigma_v$ and $\boldsymbol\sigma_w$ need to be determined. Here, a simple multivariate polynomial regression (MPR) method is used to determine the terms $\boldsymbol\tau^{(v)}_{v}(\mathbf{v})$ and $\boldsymbol\tau^{(w)}_{v}(\mathbf{v})$, where both terms are assumed to be a quadratic polynomial of $\mathbf{v}$.  The two noise coefficients $\boldsymbol\sigma_v$ and $\boldsymbol\sigma_w$ are determined by the standard deviation of the residual between the truth and the MPR fit for each shell velocity $u_n$ with $n=3,4,7$ and $8$. Therefore, both $\boldsymbol\sigma_v$ and $\boldsymbol\sigma_w$ are diagonal matrices. For a further simplification, the averaged value of the diagonal entries of $\boldsymbol\sigma_w$ is used to replace each diagonal component such that $\boldsymbol\sigma_w$ is an identity matrix times a constant, namely $\boldsymbol\sigma_w= \sigma_w\cdot\mathbf{I}$. Such a simple manipulation is preferred since the available information of the unobserved variables is very limited in practice. Sensitivity tests of the data assimilation skill with respect to such a noise coefficient $\sigma_w$ will be carried out at the end of this section.

\subsection{Skill scores}
The two skill scores adopted here to assess the data assimilation skill are the pattern correlation (Corr) and the normalized root-mean-square error (RMSE) between the reference solution (i.e., the truth) and the assimilated time series. For the conciseness of presentation, the RMSE below always stands for the normalized RMSE. The Corr and RMSE are defined as the follows,
\begin{equation}\label{SkillScores}
\begin{split}
  \mbox{Corr} &= \frac{\sum_{i=1}^I(u^{DA}_{n,i}-\bar{u}^{DA}_n)(u^{ref}_{n,i}-\bar{u}^{ref}_n)}{\sqrt{\sum_{i=1}^I(u^{DA}_{n,i}-\bar{u}^{DA}_n)^2}\sqrt{\sum_{i=1}^I(u^{ref}_{n,i}-\bar{u}^{ref}_n)^2}},\\
  \mbox{RMSE} &=  \frac{1}{\mbox{std}(u^{ref}_n)}\left(\sqrt{\frac{\sum_{i=1}^I(u^{DA}_{n,i}-u^{ref}_{n,i})^2}{I}}\right),
\end{split}
\end{equation}
where $u^{DA}_{n,i}$ and $u^{ref}_{n,i}$ are the posterior mean estimate from data assimilation and the reference solution of $u_n$, respectively, at time $t=t_i$. The value $I$ is the total number of time instants for computing these skill scores. The time averages of the forecast and the true time series are denoted by $\bar{u}^{DA}_{n}$ and $\bar{u}^{ref}_n$ while std$(u^{ref}_n)$ is the standard deviation of the reference solution.
The RMSE starts from RMSE $=0$ and loses its skill as it increases. The pattern correlation starts from Corr $=1$ and loses its skill as it decreases.

\subsection{Numerical Results}

\subsubsection{State estimation of the shell velocities associated with the unobserved variables}
\begin{figure}
\includegraphics[width=18cm]{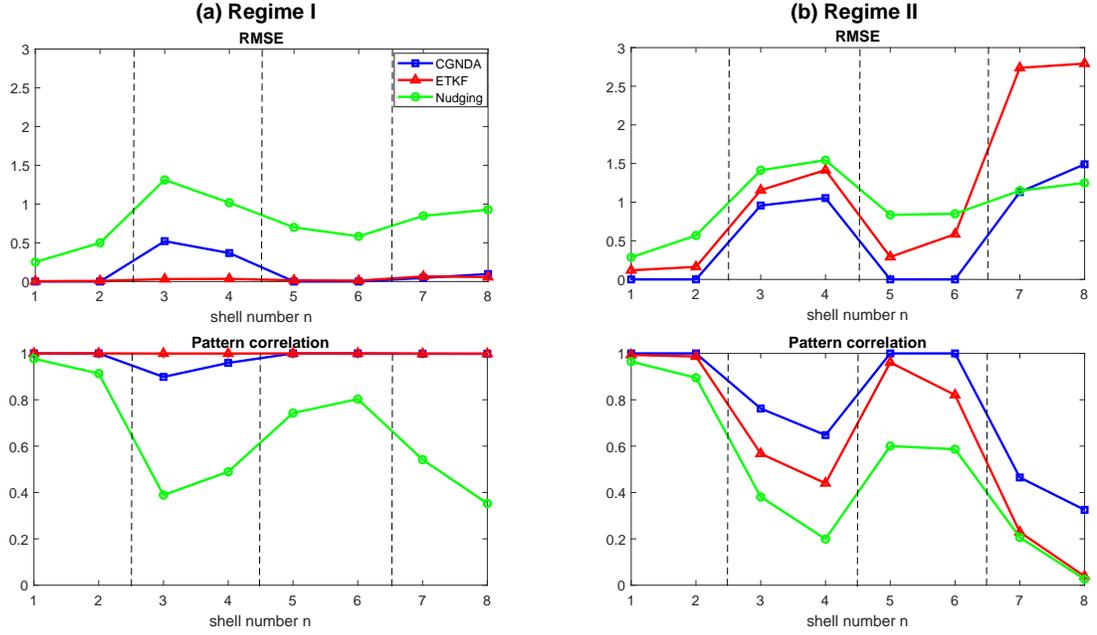}
\caption{The skill scores of the posterior mean time series from data assimilation as a function of the shell number $n$. The blue, red and green curves correspond to the CGNDA, the ETKF and the nudging data assimilation schemes. }\label{RMSE_plot_all}
\end{figure}

\begin{figure}
\includegraphics[width=18cm]{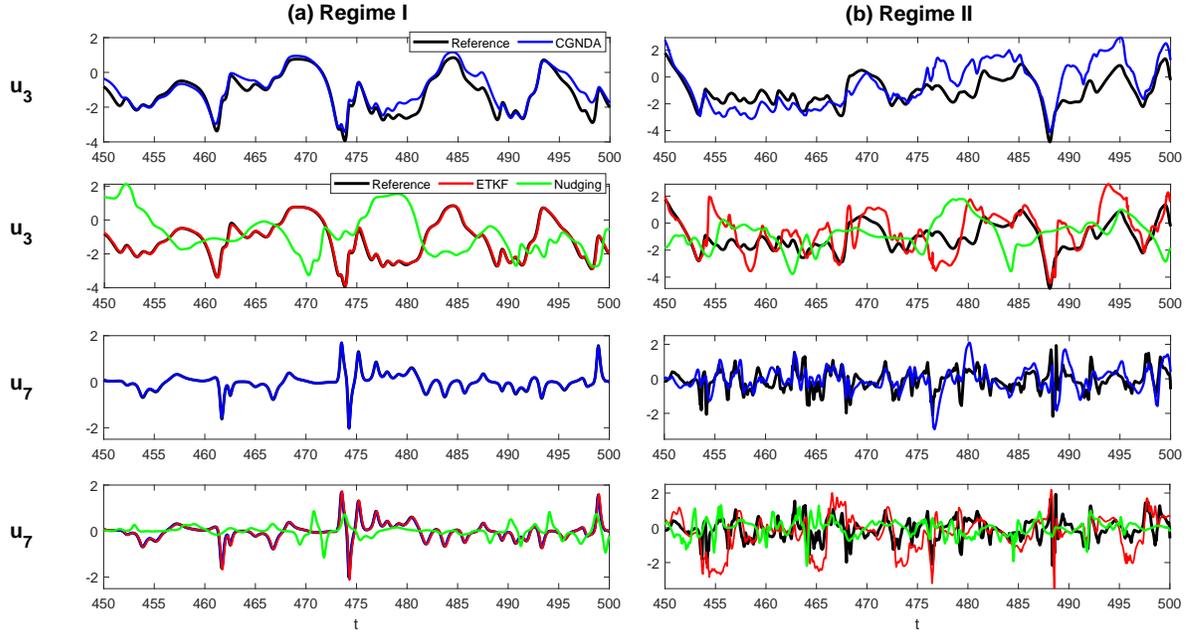}
\caption{Comparison of the posterior mean time series with the reference solution (only the real part is shown here). The black, blue, red and green curves correspond to the reference solution, the CGNDA, the ENTF and the nudging data assimilation schemes. For the convenience of comparison, the posterior mean time series from the CGNDA and those from the ETKF and nudging are shown in separate panels. The first two rows show the recovered time series of $u_3$ while the last two rows show those of $u_7$. The results of $u_4$ and $u_8$ are similar to those of $u_3$ and $u_7$, respectively. Note that the black and red curves of $u_3$ and $u_7$ as well as the black and blue curves of $u_7$ in Regime I are almost overlapping with each other. }\label{time_series_all}
\end{figure}

Figure \ref{RMSE_plot_all} shows the skill scores of the state estimation using different data assimilation methods. The focus here is on recovering the unobserved variables $u_3, u_4, u_7$ and $u_8$, although both the ETKF and the nudging schemes recover the observed variables simultaneously. The nudging scheme, even with a careful tuning of the relaxation parameter $\mu$, leads to large errors in recovering the unobserved variables as well as the observed components $u_5$ and $u_6$. This is not surprising since both dynamical regimes are chaotic. In fact, the external perturbation due to the Newtonian relaxation leads to inaccuracy in the observed variables, which passes along to the equations of the unobserved variables. As a consequence, the chaotic nature results in the biases in recovering the unobserved variables. See the green curves in Figure \ref{time_series_all} for the deviations using the nudging data assimilation scheme.

\begin{figure}
\includegraphics[width=18cm]{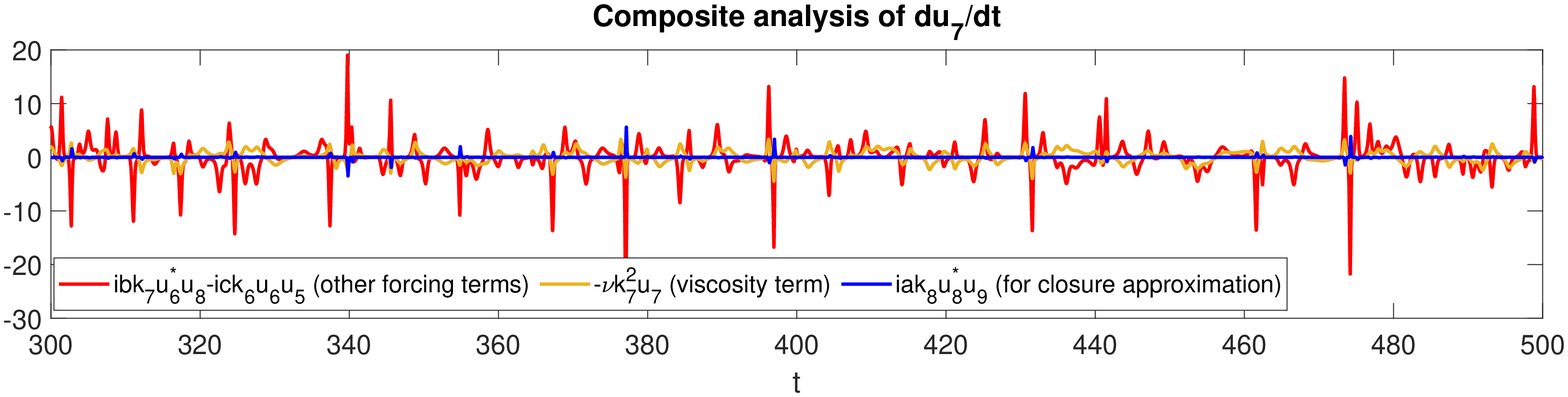}
\caption{Composite analysis of $du_7/dt$, namely the right hand side of \eqref{sabra2}. The yellow, blue and red curves correspond to $\{1\}$, $\{2\}$ and $\{3\}$ in \eqref{sabra2}. }\label{Composite_Analysis}
\end{figure}

In contrast, the CGNDA provides much more accurate results. In Regime I, although the recovered states of the medium-range scale variables $u_3$ and $u_4$ using CGNDA are not as perfect as those using the ETKF, the resulting pattern correlations are above $0.9$. The recovered trajectory of $u_3$ as shown in the first column of Figure \ref{time_series_all} matches the reference solution quite well. On the other hand, the recovered small-scale variables $u_7$ and $u_8$ using CGNDA are almost identical to the reference solution. This seems to be counter intuitive since data assimilation is in general more skillful in recovering larger scale variables. Nevertheless, $u_7$ and $u_8$ are outside the inertial range of the energy spectrum. Therefore, the dynamics of these two shells become nearly a damped-forced system. A composite analysis can be applied to the equation of $u_7$,
\begin{equation}\label{sabra2}
\frac{d u_7}{dt}
 = \underbrace{- \nu k_7^2 u_7}_{\{1\}} + \underbrace{i ak_{8} u^*_{8} u_{9}}_{\{2\}} + \underbrace{i b k_7 u^*_{6} u_{8}  - i c k_{6} u_{6} u_{5}}_{\{3\}}.
 \end{equation}
Mapping \eqref{sabra2} to the approximate model in the CGNDA \eqref{ROM_CG_b}, $\{1\}$ and $\{3\}$ in \eqref{sabra2} are retained on the right hand side of  \eqref{ROM_CG_b} while $\{2\}$ which involves higher order shells is approximated by the closure term $\boldsymbol\tau^{(w)}_{v}(\mathbf{v})$. These three terms  play different roles. The term $\{1\}$ is the ``viscous" term which damps the signal. The term $\{3\}$ does not involve $u_7$ itself, and therefore it can be regarded as an external forcing term. Similarly, the  term $\{2\}$ for the closure approximation is another external forcing. Figure \ref{Composite_Analysis} includes the composite analysis of these three components, which clearly indicates that  $\{3\}$ is the dominant one. In fact, the energy associated with $\{1\}$, $\{2\}$ and $\{3\}$ is $11.5647\%$, $0.63\%$ and $87.8052\%$, respectively. This means the approximate error is damped immediately. Different from nudging data assimilation, where the tiny error can accumulate, the estimated state is corrected by the observation in CGNDA. Such a crucial mechanism kills the error in each step and prevents the error accumulation.

The data assimilation skill using all the three methods become worse in Regime II, which is more turbulent. Nevertheless, the results from the CGNDA remains acceptable (see Figure \ref{time_series_all} and it even outweighs the ETKF, which is not suitable for nearly continuous observations and strongly turbulent systems.

It is important to note that the computational cost of the CGNDA is much lower than the ETKF. In the experiments here, the ETKF  is more than $20$ times as expensive as the CGNDA in both regimes. In addition, only an $8$-dimensional reduced order system is utilized in the CGNDA while the full perfect system is adopted for the ETKF. Nevertheless, even in the presence of model error due to the model reduction, the CGNDA overall remains as skillful as the ETKF.
 
\subsubsection{Recovery of the energy flux}
Shell models of turbulence by construction have detailed energy balance within triads of interacting waves, which is an important physical quantity to recover.
The nonlinear flux through a shell $n$ denoted by $\Pi_n$  can be computed as the difference of nonlinear transfers involving only two triads \cite{Ditlevsen2000},
\begin{equation}\label{flux}
    \Pi_n = k_{n} \Im (u^*_{n}u_{n+1}^*u_{n+2}) - (\epsilon-1) k_{n-1} \Im (u^*_{n-1}u_n^*u_{n+1})
\end{equation}
where $\Im$ denotes the imaginary part of the expression and $\epsilon = 1/2$ is used here \cite{Ditlevsen2000}.
The first term in \eqref{flux} represents the nonlinear transfer of energy from shell $n$ to shells $n+1$ and $n+2$ and the second term represents the transfer of energy from shell $n-1$ to shells $n$ and $n+1$.

\begin{figure}
\includegraphics[width=18cm]{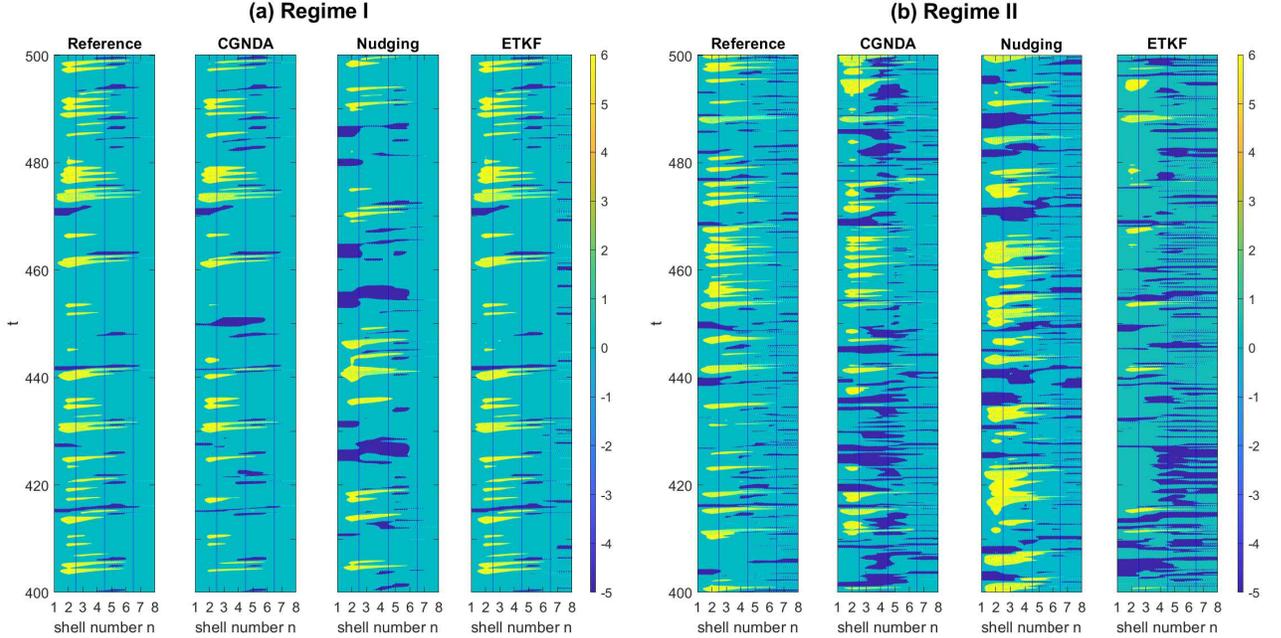}
\caption{Energy flux \eqref{flux} of the shell model. Different columns show the truth and the recovered ones using different data assimilation methods. }\label{flux_all}
\end{figure}

Figure \ref{flux_all} shows the energy flux, where shells $n$ at time $t_i$ and shells $n + 1$ at time $t_i+ \Delta t$ are connected to illustrate the energy transfer between shells.  The energy flux in shell numbers 2 to 4 obtained from the CGNDA closely resembles that of the reference solution in several different time intervals.  The intermittent bursts of energy transfer in wavenumbers 6 to 8  for CGNDA also have a nearly perfect  match with the arrival of these bursts in the reference solution.

\subsubsection{Closure approximations}
\begin{figure}
\includegraphics[width=18cm]{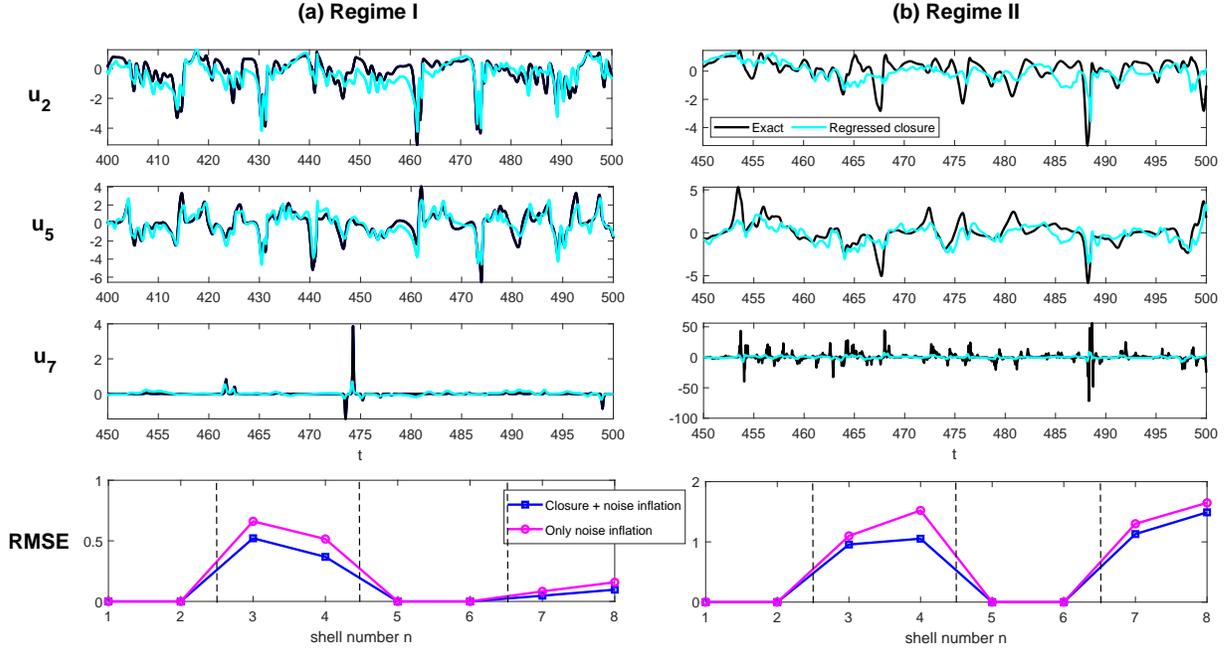}
\caption{Comparison of the left hand side of \eqref{Parameterization_Terms} (black; named as ``exact'') and the right hand side closure term $\boldsymbol\tau^{(v)}_{v}(\mathbf{v})$ or $\boldsymbol\tau^{(w)}_{v}(\mathbf{v})$ based on the quadratic approximation  (cyan; named as ``regressed closure''). The last row shows the RMSE of the CGNDA using the closure approximation \eqref{Parameterization_Terms} (blue) and the one with only noise inflation but without the closure terms  $\boldsymbol\tau^{(v)}_{v}(\mathbf{v})$ and $\boldsymbol\tau^{(w)}_{v}(\mathbf{v})$ (magenta).}\label{regression_all}
\end{figure}

Recall that a closure approximation \eqref{Parameterization_Terms} is adopted in the CGNDA. It is therefore important to understand the accuracy in applying such a closure approximation. Figure \ref{regression_all} compares the left hand side of \eqref{Parameterization_Terms} and the closure term $\boldsymbol\tau^{(v)}_{v}(\mathbf{v})$ or $\boldsymbol\tau^{(w)}_{v}(\mathbf{v})$ based on the MPR with a quadratic approximation on the right hand side. For the observed variables $u_2$ and $u_5$, the closure illustrates a high skill in approximating the truth. The pattern correlation between the closure approximation and the truth is above $0.8$ for both $u_2$ and $u_5$ in Regime I and that is above $0.5$ in Regime II. In addition, the intermittent events are captured quite accurately in both the regimes.

One natural question is how the closure approximation \eqref{Parameterization_Terms} improves the data assimilation skill. To this end, the data assimilation skill of the CGNDA with only the noise inflation but without the closure terms  $\boldsymbol\tau^{(v)}_{v}(\mathbf{v})$ and $\boldsymbol\tau^{(w)}_{v}(\mathbf{v})$ is carried out. The comparison between this cruder approximation with the closure approximation is shown in the last row of Figure \ref{regression_all}. Note that the noise coefficients in these two methods are different. In the cruder method with only noise inflation, the noise amplitude is based on the entire variance on the left hand side of \eqref{Parameterization_Terms}. Noise inflation has been shown to improve the data assimilation skill \cite{anderson2007adaptive, whitaker2012evaluating} and therefore this method remains reasonable results. Yet, comparing with the CGNDA with the closure approximation \eqref{Parameterization_Terms}, it is obvious that using only the noise inflation is not as skillful as the closure approximation, which indicates the necessity of the closure terms.

\subsubsection{Suboptimal data assimilation and sensitivity analysis}

\begin{figure}
\includegraphics[width=18cm]{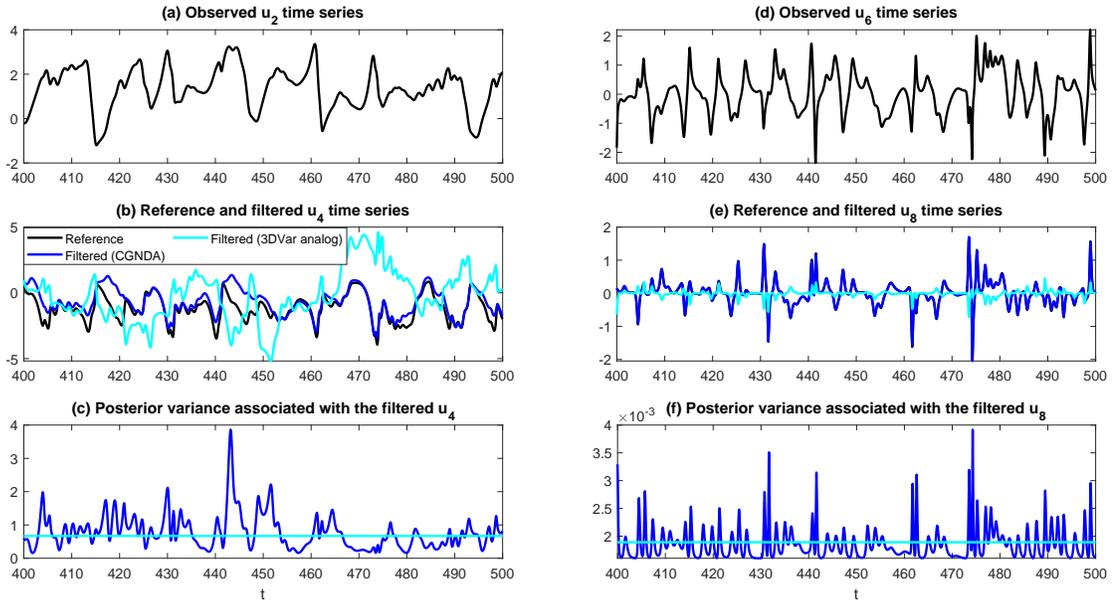}
\caption{The effect of the time evolution of the posterior covariance in Regime I. The blue curves are those computed from the CGNDA posterior update formulae \eqref{CGNS_Stat} while the cyan ones are from using only the posterior mean update \eqref{CGNS_Stat_Mean} while the posterior covariance is fixed. Therefore, the latter is named as a 3DVar analog. Note that in Panel (e) the black and blue curves are almost overlapped with each other.}\label{3DVar}
\end{figure}

Different from the nudging data assimilation, the CGNDA has one advantage that the posterior covariance is a time-dependent function. Therefore, it is important to understand the role of such a non-stationary covariance matrix in facilitating the data assimilation skill. Figure \ref{3DVar} compares the data assimilation skill with such a time-dependent posterior covariance and that with a constant one for Regime I. From Panels (c) and (f), it is clear that the posterior covariance (blue) varies significantly in time. The peaks of the posterior covariance align well with the extreme events in the unobserved time series (especially for $u_8$), which indicates the necessity in recovering the intermittent features using such a non-stationary uncertainty evolution.
As an analog to the 3DVar algorithm \cite{kalnay2003atmospheric}, we set the posterior covariance to be a constant. Here, the time averaged value of each entry of $\mathbf{R}$ is first used to form an auxiliary matrix $\tilde{\mathbf{R}}$. Then an eigenvalue decomposition is applied to obtain $\tilde{\mathbf{R}}\tilde{\mathbf{R}}^* = \mathbf{B}\boldsymbol{\Lambda}\mathbf{B}^*$. The constant posterior covariance is then given by $\tilde{\mathbf{R}}_C = \mathbf{B}\sqrt{\boldsymbol\Lambda}\mathbf{B}^*$. In Panels (c) and (f) of Figure \ref{3DVar}, the cyan curves show the diagonal components of such a constant posterior covariance corresponding to $u_4$ and $u_8$. Plugging $\tilde{\mathbf{R}}_C$ into \eqref{CGNS_Stat_Mean}, the resulting posterior mean time series from such a 3DVar analog filter is shown in the cyan curves in Panels (b) and (e), which are much worse than using the exact posterior update formulae \eqref{CGNS_Stat} of the CGNDA. On the other hand, if a constant posterior covariance is built in the above way, then the corresponding posterior mean time series blows up in Regime II. Therefore, the time evolution of the posterior covariance, representing the filter uncertainty, is extremely important in recovering these turbulent flows.

Another sensitivity test has been done for the values of $\sigma_w$ in Section \ref{Sec:coefficients}. It has been shown that the RMSE differs by only $20\%$ when $\sigma_w$ varies around the value used here. Note that the criterion developed in Section \ref{Sec:coefficients} leads to a $\sigma_w$ that is not the optimal value, which is slightly smaller. But the simple criterion is easy to implement in practice. Note that if $\sigma_w$ is too small, then the data assimilation scheme blows up, which is consistent with many existing filtering results \cite{harlim2010catastrophic, willems1992divergence, fitzgerald1971divergence}.

\section{Conclusion}\label{Sec:Conclusion}
In this article an efficient continuous in time data assimilation scheme called the CGNDA is developed. A simple approximation modeling framework is utilized to transform a complex nonlinear system to fit the  required mathematical structure of the CGNDA scheme. The new algorithm is applied to the Sabra shell model, which is a conceptual model for turbulence. It has been shown that the CGNDA outweighs both the ETKF and the nudging data assimilation schemes in terms of both the accuracy and computational efficiency. 
As a future work, the framework \eqref{ROM_CG2} with a more refined reduced order model structure can be incorporated with the CGNDA for the ensemble forecast. It is interesting to see how the refined structure, together with the physics constraints, improves the forecast skill.

\section{Acknowledgments}
The research of N.C. is partially funded by the Office of VCRGE at UW-Madison.  Y. L.  is supported as a graduate student under the VCRGE grant.  E.L is thankful for the kind hospitality of University of Wisconsin-Madison, Department of Mathematics, where part of this work was completed.   The work of E.L. is supported by the ONR grant N0001421WX01362. The data that support the findings of this study are available from the corresponding author upon reasonable request.


%
%

%



\end{document}